\documentclass[superscriptaddress,aps,prl,floats,onecolumn,twoside,floatfix]{revtex4}
\usepackage{epsfig}
\usepackage{amsmath}
\usepackage{amssymb}
\usepackage[polutonikogreek,english]{babel}
\usepackage[or]{teubner}

\newcommand{\be}{\begin{equation}}
\newcommand{\ee}{\end{equation}}
\newcommand{\bea}{\begin{eqnarray}}
\newcommand{\eea}{\end{eqnarray}}
\renewcommand{\a}{\alpha}

\begin{document}

\title{Ultrametricity in the Edwards-Anderson Model}  % Enter your title between curly braces

\author{Pierluigi Contucci}
\affiliation{
%Dipartimento di Matematica,
Universit\`{a} di Bologna, Piazza di Porta S.Donato 5, 40127 Bologna, Italy}

\author{Cristian Giardin\`a}
\affiliation{Eurandom, P.O. Box 513 - 5600 MB Eindhoven, The Netherlands}

\author{Claudio Giberti}
\affiliation{ Universit\`a di Modena e Reggio Emilia, via G.
Amendola 2 -Pad. Morselli- 42100 Reggio Emilia, Italy}

\author{Giorgio Parisi}
\affiliation{
Universit\`a La Sapienza di Roma,  Roma, Italy}

\author{Cecilia Vernia}
\affiliation{
Universit\`{a} di Modena e Reggio Emilia, via Campi 213/B, 41100 Modena, Italy}

\begin{abstract}
We test the property of ultrametricity for the spin glass three-dimensional Edwards-Anderson model in zero
magnetic field with numerical simulations up to $20^3$ spins.
We find an excellent agreement with the prediction of the mean field theory. Since ultrametricity is not
compatible with a trivial structure of the overlap distribution our result
contradicts the droplet theory.
\end{abstract}

\maketitle

\rightline{
\textgreek{$^{\r{}}$O \As anax o\Cr u t\`o mante\~i\'on \s esti t\`o \s en Delfo\~is}
}
\rightline{
\textgreek{o\As ute l\'egei o\As ute kr\'uptei \s all\` a shma\'inei}
}
\rightline{\it Heraclitus Fragment 93,}
\rightline{from Plutarch, On the Pythian Oracle, 404E. \cite{KRS}}

\vskip 1truecm

Ultrametricity is a widely accepted property of the mean field spin glass theory: it is a crucial
ingredient in the field theoretical computations of the Sherrington-Kirkpatrick model
\cite{SK,MPV,Pa} as well as a guiding principle for the rigorous proof of its free energy density formula \cite{G1,T}.
Its relevance in finite dimensional systems is nonetheless still an open matter, subject of
intense investigations and debates in the theoretical and mathematical
physics communities.

Ultrametricity states a very striking property
for a physical system: essentially it says that the equilibrium configurations of a large system
can be classified in a taxonomic (hierarchical) way (as animal in different taxa): configurations are grouped in
states, states are grouped in families, families are grouped in superfamilies. This equilibrium
ultrametricity has a correspondence in the existence of widely separated time scales in the
dynamics, typically of a glassy system.

It is not clear at the present moment if ultrametricity is present in three dimensional systems; the
most studied case is three dimensional spin glasses where contrasting results have been presented in
the literature in the last twenty years. Part of the difficulties arise from the fact that
ultrametricity should be, at the best, exact when the volume of the system goes to infinity and
therefore simulations done on a limited range of volume are difficult to interpret. In this letter
we study systems ranging from $4^{3}$  to $20^{3}$ extending of about an order of magnitude the
range of  volume used in previous simulations.

From the technical point of view ultrametricity implies that sampling three configurations independently with
respect to their common Boltzmann-Gibbs state and averaging over the disorder, the distribution of the
distances among them is supported, in the limit of very large systems, only on equilateral and isosceles
triangles with no scalene triangles contribution. In a generic situation the relative weight of equilateral
and isosceles triangles is arbitrary, however it is well established in \textit{stochastically stable} systems;
the {\it stochastic stability} property was introduced for the infinite range spin glass model in \cite{AC,G}
and later proved also for the realistic short ranged models in finite dimensions \cite{C,CG}.

The property of ultrametricity and the non-trivial structure of the overlap distribution
are the characterizing features of the mean field picture and are mutually intertwined:
a trivial (delta-like) overlap probability distribution, like the one predicted in the droplet theory
\cite{FH}, is not compatible in fact with the previous
ultrametric structure because it predicts only equilateral triangles all of the same side.

In this letter we study the Edwards-Anderson model \cite{EA} for the spin glass in the
three-dimensional cubic lattice with $\pm J$ random interactions (for a numerical study in four dimensions see \cite{CMP}).
With a multi-spin coding and a parallel-tempering algorithm we numerically investigate the
distribution of the overlaps: all the parameters used in the simulations are reported in Tab.\ref{t:para}.
We have checked the thermalization by verifying that our result would have been the same (inside our small
error bar) by taking simulations a factor 4 shorter.
\begin{table}
\begin{center}
\begin{tabular}{|c|c|c|c|c|c|}\hline
$L$   & Sweeps   & Nreal  & $n_{\beta}$ & $T_{min}$ & $T_{max}$\\ \hline \hline
$4$  & $1047552$ & $1280$ &   $25$   &  $0.7$    &   $2.1$ \\ \hline
$6$  & $1047552$ & $1280$ &   $25$   &  $0.7$    &   $2.1$ \\ \hline
$8$  & $1047552$ & $1280$ &   $25$   &  $0.7$    &   $2.1$ \\ \hline
$10$  & $1047552$ & $1280$ &   $25$   &  $0.7$    &   $2.1$ \\ \hline
$12$  & $1047552$ & $896$ &   $25$    &  $0.7$    &   $2.1$ \\ \hline
$16$  & $2096128$ & $1216$ &   $25$  &  $0.7$    &   $2.1$ \\ \hline
$18$  & $2096128$ & $768$ &   $49$   &  $0.7$    &   $2.1$ \\ \hline
$20$  & $4193280$ & $512$ &   $103$  &  $0.7$    &   $2.1$ \\ \hline
\end{tabular}\caption{Parameters of the simulations: system size, number of sweeps, number
of disorder realizations, number of temperature values allowed in the parallel tempering procedure,
minimum and maximum temperature values.}\label{t:para}
\end{center}
\end{table}

We find very strong indication in favor of ultrametricity which
turns out to be reached at large volumes with exactly the form
predicted by the mean field theory and, by consequence, a robust
signal against droplet theory (for a study of dynamical
ultrametricity and for the relation between statics and dynamics in
spin glasses see \cite{FRT,FMPP}). According to the literature the
system has a transition $T_{c}$ of 1.15 and our data are compatible
with this value. The smallest temperature we used is 0.7, i.e. about
$0.6T_{c}$: although we are relatively far from the critical
temperature, we may still feel some effects coming from the critical
region. However we notice that ultrametricity should not be valid at
the critical temperature, consistent with our results, so any
finding of ultrametricity at lower temperature cannot be an artifact.

From a mathematical point of view the triple $(c_{1,2},c_{2,3},c_{3,1})$, with $0 \le c_{i,j} \le 1$, representing the
overlaps among three copies of the system,
is called {\it stochastically stable and ultrametric} when, defining $\chi(c)=\int_{0}^{c}P(c')dc'$, where
$P(c)$ is the probability distribution of $c$, its
joint probability distribution function has the following structure:
\begin{equation}\label{um}
P_{3}(c_{1,2},c_{2,3},c_{3,1}) = \frac12P(c_{1,2})\chi(c_{1,2})\delta(c_{1,2}-c_{2,3})\delta(c_{2,3}-c_{3,1})
\end{equation}
\begin{equation}\nonumber
+
\frac12P(c_{1,2})P(c_{2,3})\theta(c_{1,2}-c_{2,3})\delta(c_{2,3}-c_{3,1})
\end{equation}
\begin{equation}\nonumber
+
\frac12P(c_{2,3})P(c_{3,1})\theta(c_{2,3}-c_{3,1})\delta(c_{3,1}-c_{1,2})
\end{equation}
\begin{equation}\nonumber
+
\frac12P(c_{3,1})P(c_{1,2})\theta(c_{3,1}-c_{1,2})\delta(c_{1,2}-c_{2,3}) \; .
\end{equation}

Thinking of the quantities $c$'s as $1$ minus the sides of a triangle the previous formula says that only
equilateral (first term on the right hand side of eq.  (\ref{um})) and isosceles (last three terms of eq.
(\ref{um})) triangles are allowed, the scalene triangles have zero probability.  Equation (\ref{um}) implies
that the
distribution of the three random variables $u=\min(c_{1,2},c_{2,3},c_{3,1})$,
$v=\mbox{med}(c_{1,2},c_{2,3},c_{3,1})$ and $z=\max(c_{1,2},c_{2,3},c_{3,1})$ is
\begin{equation}\label{ro}
\rho(u,v,z)=\frac12\chi(u)P(u)\delta(v-u)\delta(z-v)+\frac32 P(z)P(v)\theta(z-v)\delta(v-u) \; ,
\end{equation}
and from that one deduces that the distribution of the two differences $x=v-u$, $y=z-v$ is
\begin{equation}
\tilde\rho(x,y)=\delta(x)\left[\frac14\delta(y)+\frac32\theta(y)\int_y^1 P(a)P(a-y)da\right ] \; ,
\end{equation}
whose marginals are
\begin{equation}\label{medmin}
\tilde\rho(x)=\delta(x) \; ,
\end{equation}
\begin{equation}\label{maxmed}
\tilde\rho(y)=\frac14\delta(y)+\frac32\theta(y)\int_y^1 P(a)P(a-y) da \; .
\end{equation}
We recall that the Hamiltonian of the EA model \cite{EA} is given by
\begin{equation}
H_\sigma=-\sum_{|i-j|=1}J_{i,j}\sigma_i\sigma_j
\end{equation}
with $J_{i,j}=\pm 1$ symmetrically distributed and Ising spins $\sigma_i = \pm 1$. Given two spin configurations
$\sigma$ and $\tau$
for a system of linear size $L$, we consider the main observables: the link-overlap
\be
Q(\sigma,\tau)=(3L^3)^{-1}\sum_{|i-j|=1}\sigma_i\sigma_j\tau_i\tau_j
\ee
which is the normalized Hamiltonian covariance, and the standard overlap
\be
q(\sigma,\tau)=(L^3)^{-1}\sum_{i}\sigma_i\tau_i
\ee
which is related to the Edwards-Anderson order parameter. For every function of two spin configurations
$c(\sigma,\tau)$ (for instance $Q$ or $q$) the physical model induces a probability distribution by the formula
\begin{equation}\label{pc}
{\cal P}_3(c_{1,2},c_{2,3},c_{3,1}) =
\langle
%\sum_{\sigma,\tau,\gamma}
\delta(c_{1,2}-c(\sigma,\tau))\delta(c_{2,3}-c(\tau,\gamma))\delta(c_{3,1}-c(\gamma,\sigma))
\rangle \; ,
\end{equation}
where $\sigma,\tau,\gamma$ denote three different equilibrium configurations.  Here and in the sequel the
brackets $\langle \cdot \rangle$ will denote the average over the disorder $J_{i,j}$ of the thermal average
over the Boltzmann-Gibbs distribution.

We will find very strong evidences that for large volumes the link overlap
has the ultrametric structure of eq. (\ref{um}).  (we are in zero magnetic field and the system is invariant
under a global change of all the spins). As we shall see at the end of the paper the same results are valid
also for the standard overlap with the only difference that it has a symmetric distribution in the interval
$[-1,1]$ and the triangle distribution is built on it by suitable contributions of the positive and negative
values, see formula (\ref{gium}) below.

We present firstly the results for the link overlap for two reasons: the analysis
is conceptually simpler, the link overlap is more fundamental than the standard overlap and contains more
interesting information, e.g. two configurations that differ by a spin inversion of a compact region of size
half of the lattice, will have, in the infinite volume limit, a zero standard overlap, but a large link overlap.

The results can be described as follows. We test numerically the structure of the distribution
for the two random variables $X=Q_{med}-Q_{min}$ and $Y=Q_{max}-Q_{med}$ where the $Q$'s
represent the largest, medium, and smaller value of the link-overlap
among three copies of the system. The numerical data are compared to the formulas (\ref{medmin})
and (\ref{maxmed}).

\begin{itemize}
%\item Figure 1:
%The two panels show the plot of $<X>/<Q>$ (left) and $<Y>/<Q>$ (right) as functions of the temperature $T$.
%We see that the average value of the $X$ variable is much closer to zero than that
%of the $Y$ variable. Nevertheless both averaged variables show some tendency toward
%zero. Since ultrametricity implies that the first quantity goes to zero while the other doesn't
%we perform the analysis of the second moments and of the distributions in order to
%resolve the behavior of the two variables.

\item Figure 1:
We find that the variances of the two variables have a totally different behavior. The
left panel contains the plot of $Var(X)/Var(Q)$ and the right panel of $Var(Y)/Var(Q)$
both as a function of $Var(Q)$. We find more convenient this parametrization with respect
to the usual one using temperature because it allows to extract more information on
size dependence through scaling laws: this is due to the fact that $Var(X)$,$Var(Y)$ and $Var(Q)$
have size dependence changing with $T$. In particular within the temperature range that
we have taken into account the quantity $Var(Q)$ decreases monotonically with the temperature.
The figure clearly shows that while the variance of $X$ is shrinking to zero
the variance of $Y$ is growing with the volume. Moreover the variance of $X$ satisfy
a scaling law with very good accuracy: $Var(X)/Var(Q)$ scales like $L^{-1.18}$ (see inset)
while there is no scaling law for the second variable.

\item Figure 2:
The figure displays for two system sizes of $L=12$ and $L=20$ the data histograms
for $X$ (in black) and $Y$ (in red) variable at $T=0.7$.
They show that ${\cal P}(X)$, the empirical distribution of $X$, is much more concentrated close to zero,
while ${\cal P}(Y)$ is spread on a
larger scale. The function $\tilde \rho(Y)$ provides a test of consistency with formula (\ref{maxmed}).
The plot of $\tilde\rho(Y)$ has been obtained using the data histograms of $X$ to represent
the delta function (\ref{medmin}) and the experimental data for the distribution of $Q$ inside
the convolution. The two curves superimpose each other with an excellent agreement. We have
also tested that any different numerical weight other than $1/4$ and $3/4$ do not yield
such an agreement.
\end{itemize}

The previous results clearly show that the link overlap has an ultrametric distribution.
Our next investigation is about the standard overlap for which we find that it also
obeys ultrametricity. Given the three standard overlaps $q_{1,2},q_{2,3},q_{3,1}$ their
probability measure is a priori supported on $[-1,1]^3$. Reflection invariance
($q_{i,j}\to \alpha_i q_{i,j} \alpha_j$, with $\alpha=\pm 1$) implies that
it is a sum of two orbits, one for $S=\mbox{sign} (q_{1,2}q_{2,3}q_{3,1})>0$ and the other
for $S<0$. The mean field theory predicts that only non-frustrated triples ($S>0$) contribute
to the triangle distribution, namely:
\be\label{gium}
{\bar P}_3(q_{1,2},q_{2,3},q_{3,1})=\frac{1}{4}\left[P_3(q_{1,2},q_{2,3},q_{3,1})\theta(q_{1,2})\theta(q_{2,3})\theta(q_{3,1})
+P_3(-q_{1,2},-q_{2,3},q_{3,1})\theta(-q_{1,2})\theta(-q_{2,3})\theta(q_{3,1})\right.
\ee
\be\nonumber
\left. +P_3(q_{1,2},-q_{2,3},-q_{3,1})\theta(q_{1,2})\theta(-q_{2,3})\theta(-q_{3,1})
+P_3(-q_{1,2},q_{2,3},-q_{3,1})\theta(-q_{1,2})\theta(q_{2,3})\theta(-q_{3,1})
 \right]
\ee
To check the validity of the previous formula it is convenient to introduce
the new random variables
\be
\tilde q_{max} = \max (|q_{1,2}|,|q_{2,3}|,|q_{3,1}|)
\ee
\be
\tilde q_{med} = \mbox{med} (|q_{1,2}|,|q_{2,3}|,|q_{3,1}|)
\ee
\be
\tilde q_{min} = \mbox{sign} (q_{1,2}q_{2,3}q_{3,1}) \min (|q_{1,2}|,|q_{2,3}|,|q_{3,1}|)
\ee
and verify that their distribution is the (\ref{um}).
The numerical results are illustrated in Fig 3: the left panel shows
how the normalized variance of the variable $\tilde x=\tilde q_{med}-\tilde q_{min}$ has a clear
tendency to vanish for temperatures below the critical point. The inset displays the log-log plot
of $Var(\tilde x)/Var(|q|)$ as a function of $L$ at the lowest available temperature $T=0.7$.
At the critical point the quantity is instead size invariant as predicted
by the mean field theory. A totally different behavior is found for the variable
$\tilde y=\tilde q_{max}-\tilde q_{med}$ where below the critical temperature the normalized variance
is increasing but still size invariant at criticality.

We have also explicitly investigated the contribution of the frustrated triples by plotting the quantity
$S^{(-)} = \int_{-1}^{0} d\tilde q_{min} p(\tilde q_{min})\tilde q^2_{min}/\int_{-1}^{1} d\tilde
q_{min}p(\tilde q_{min})\tilde q^2_{min}$, where $p(\tilde q_{min})$ is the probability distribution of
$\tilde q_{min}$: the left panel of Fig. 4 clearly shows that the distribution
of $\tilde q_{min}$ is supported
almost completely on the positive interval and that the negative values are concentrated
near zero (for similar quantities and other three-replicas observables see \cite{IPRL}).
This implies that the contribution associated to
the frustrated orbit ($S<0$) is very small at large volumes.

The equivalent behavior of link and standard overlap is indeed expected because it extends
previous findings of \cite{CGGV,MaPa} where it was shown
that link and standard overlaps are mutually non fluctuating for the case of Gaussian couplings.
In the right panel of Fig. 4  we show, for the model with $\pm J$ investigated within this
work, the analysis of the relative fluctuation and functional dependence of the two overlaps.
It is shown the function $G(q^2)=\langle Q|q^2\rangle$, i.e. the expected value of the link-overlap for
an assigned value of the standard overlap, for different system sizes at $T=0.7$,
with a fit to the infinite volume limit $g_\infty(q^2)$.
The conditional variance of $Q$ given $q^2$, displayed in the inset, shows a
trend toward a vanishing value for infinite system sizes.

Numerical simulations, like the Delphi Oracle for Heraclitus, neither conceal nor reveal
the truth, but only hint at it. In this work we have investigated the property of
ultrametricity in a short-range spin-glass model.
We have shown that violations of ultrametricity in finite volumes have
a clear tendency to vanish as the system size increases.
We verified moreover that the analytical predictions of the ultrametric
replica symmetry breaking ansatz are correct up to the tested sizes.
Our results contradict previous finding \cite{HYD} done for much smaller volumes (up to $8^3$)
in which lack of ultrametricity was claimed. We have shown instead strong numerical evidence
that the spin glass in three dimensions fulfills the property of ultrametricity for both the link and the
standard overlap distributions. A detailed account of the present investigation will appear elsewhere \cite{CGGPV}.

\vskip 1cm

{\bf Acknowledgments.} We thank S. Graffi, F. Guerra, E. Marinari, C. Newman, D. Stein and F. Zuliani
for useful discussions. C. Giardin\`a and C. Vernia acknowledge GNFM-INdAM for financial support.

\begin{figure}
    \setlength{\unitlength}{1cm}
          \centering
               \includegraphics[width=11cm,height=6cm]{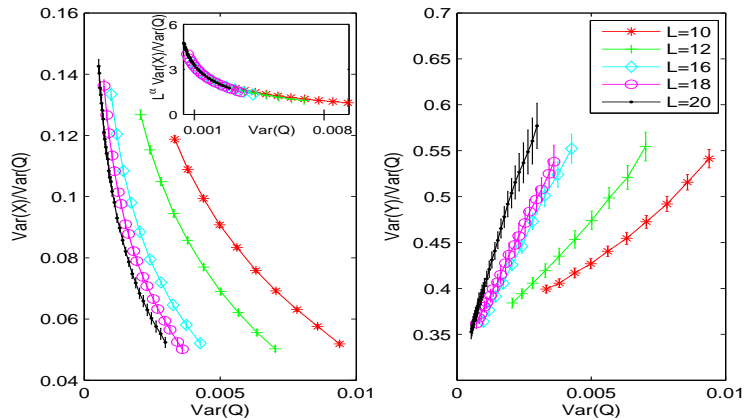}
               \caption{Normalized variances of the two random variables
$X=Q_{med}-Q_{min}$ (left) and $Y=Q_{max}-Q_{med}$ (right) as a function of $Var(Q)$. The
inset (at left) shows the scaling law for $\a=1.18$, i.e. $L^\alpha
Var(X)/Var(Q)$ is $L$-independent.}

\label{fig1}
\end{figure}

\begin{figure}
    \setlength{\unitlength}{1cm}
          \centering
               \includegraphics[width=11cm,height=6cm]{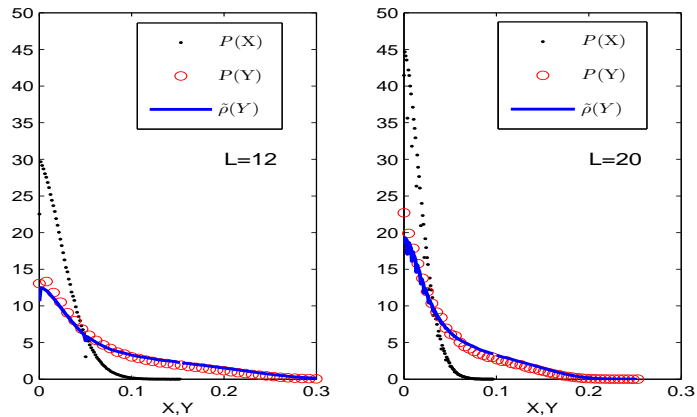}
               \caption{Empirical distributions ${\cal P}(X)$ and ${\cal P}(Y)$ for $X=Q_{med}-Q_{min}$
               and $Y=Q_{max}-Q_{med}$ for the two system sizes ($L=12$ and
               $L=20$) at temperature $T=0.7$. $\tilde\rho(Y)$ shows the distribution of $Y$
               computed from formula (\ref{maxmed}) using experimental data for ${\cal P}(Q)$
               and approximating the delta function with the histogram of $X$}
\label{fig2}
\end{figure}
%{Fig4PRL.eps}
\begin{figure}
    \setlength{\unitlength}{1cm}
          \centering
               \includegraphics[width=11cm,height=6cm]{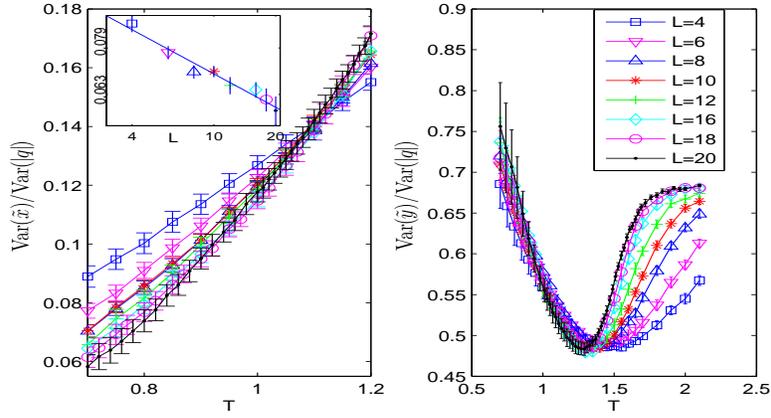}
               \caption{Normalized variances of the two random variables
$\tilde x=\tilde q_{med}- \tilde q_{min}$ (left) and $\tilde y=\tilde q_{max}- \tilde q_{med}$ (right) as
a function of the temperature. The inset (at left) shows the $L$-dependence of $Var(\tilde x)/Var(|q|)$ at fixed
temperature $T=0.7$ on a log-log scale together with the best fit: $Var(\tilde x)/Var(|q|)\sim aL^b$, $a=0.12, b=-0.23$.}

\label{fig3}
\end{figure}

\begin{figure}
    \setlength{\unitlength}{1cm}
          \centering
               \includegraphics[width=11cm,height=6cm]{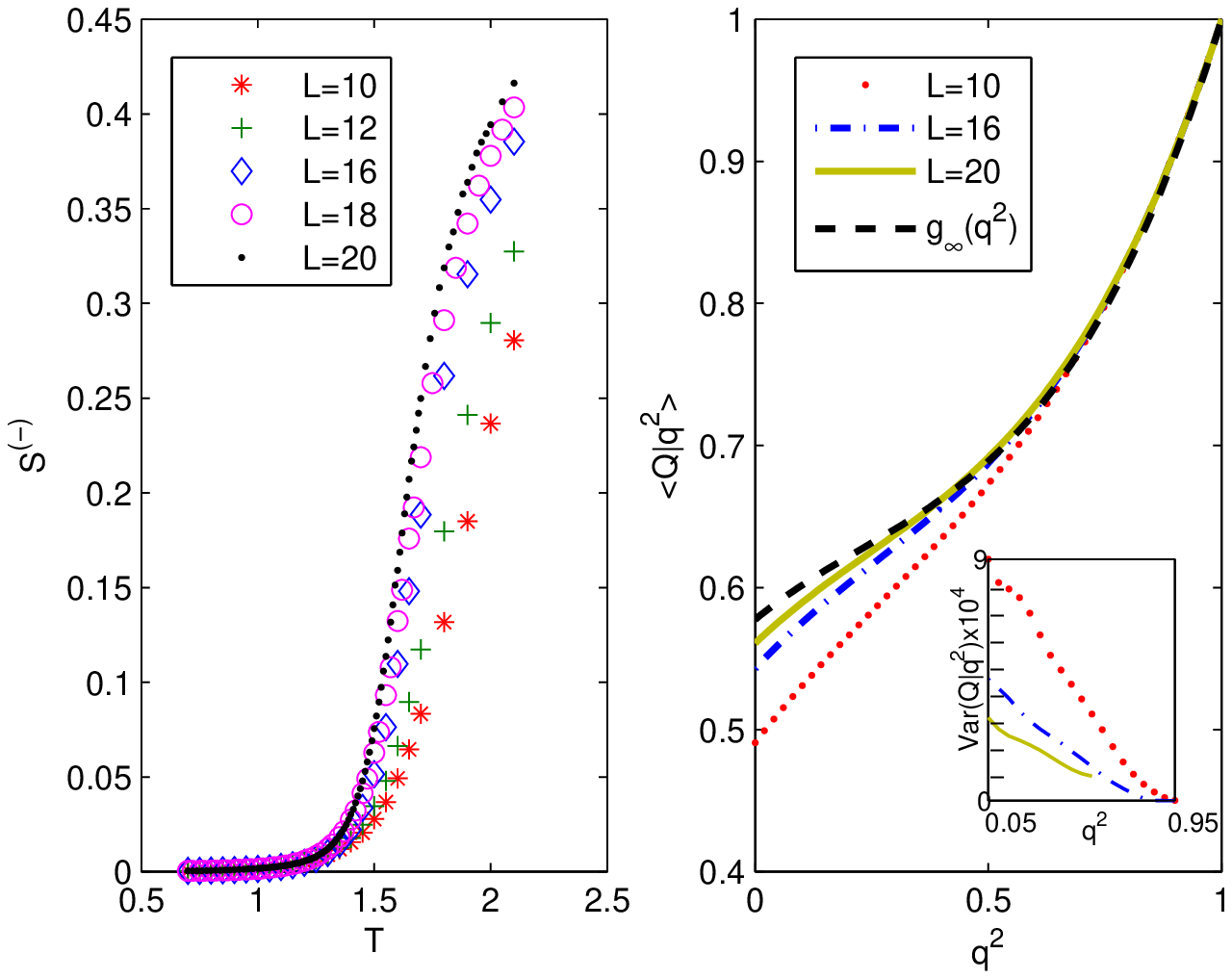}
               \caption{Left panel: the average value of $S^{(-)}$ (defined in the text)
               as a function of $T$.
               Right panel: Conditional expectation $\langle Q|q^2\rangle$ and conditional
               variance $Var(Q|q^2)$(inset) of the random variable
               $Q$ given $q^2$, where $Q$ is the link-overlap and $q^2$ is the square
               of the standard overlap, for different system sizes at temperature $T=0.7$}
\label{fig4}
\end{figure}


\begin{thebibliography}{99}

\bibitem{KRS} G.S.Kirk, J.E.Raven,M.Schofield.,
{\em The Presocratic Philosophers}
Cambridge University Press (1995)

\bibitem{SK} D. Sherrington and S. Kirkpatrick,
%``Solvable Model of a Spin Glass'',
{\em Phys. Rev. Lett.} {\bf 35}, 1792 (1975).

\bibitem{MPV}
M. Mezard, G. Parisi, M.A. Virasoro,
{\em Spin Glass Theory and Beyond}
World Scientific, Singapore (1987).

\bibitem{Pa} G. Parisi, F. Ricci-Tersenghi,
%``On the origin of ultrametricity'',
{\em J. Phys. A: Math. Gen.} {\bf 33}, 113, (2000).

\bibitem{G1} F. Guerra,
% replica symmetry breaking
{\em Comm. Math. Phys.} {\bf 233}, 1, (2003).

\bibitem{T} M. Talagrand,
% parisi proof
{\em Annals of Math.} {\bf 163}, 221 (2006).
This result is announced in {\em C.R.A.S.} {\bf 337}, 111, (2003).

\bibitem{AC} M. Aizenman, P. Contucci
% stochastic stability
{\em J. Stat. Phys.} {\bf 92}, 765, (1998).

\bibitem{G} F. Guerra,
% about the overlap distribution
{\em Int. Jou. Phys. B} {\bf 10}, 1675, (1997).

\bibitem{C} P. Contucci,
%``Replica Equivalence in the Edwards-Anderson Model''
{\em J. Phys. A: Math. Gen.} {\bf 36}, 10961, (2003).

\bibitem{CG} P. Contucci, C. Giardin\`a,
%``the Ghirlanda Guerra identitites'',
{\em Jour. Stat. Phys.} to appear (2006).
{\em math-ph/0505055}

\bibitem{FH} D.S. Fisher and D.A. Huse,
{\em Phys. Rev. Lett.} {\bf 56}, 1601 (1986)

\bibitem{EA} S.F. Edwards and P.W. Anderson,
%``Theory of spin glasses''
{\em Jou. Phys. F.}, {\bf 5}, 965, (1975).


\bibitem{CMP} A. Cacciuto, E. Marinari, G. Parisi
{\em J. Phys. A: Math. Gen} {\bf 30} L263-L269 (1997)
%"A numerical study of ultrametricity in finite-dimensional spin glasses"
%per l'ultrametricita' in quattro dimensioni.


\bibitem{FRT} S. Franz, F. Ricci-Tersenghi
{\em Phys. Rev. E} {\bf 61}, 1121-1124 (2000)
%per uno studio  dinamico dell'ultrametricita' in tre dimensioni
%"Ultrametricity in three-dimensional Edwards-Anderson spin glasses"

\bibitem{FMPP} S. Franz, M.Mezard, G.Parisi, L.Peliti
{\em Phys. Rev. Lett.} {\bf 81}, 1758 (1998)
%per uno studio  dinamico dell'ultrametricita' in tre dimensioni
%"Ultrametricity in three-dimensional Edwards-Anderson spin glasses"

\bibitem{IPRL} D. I\~{n}iguez, G. Parisi, J. Ruiz-Lorenzo
{\em J. Phys. A: Math. Gen.} {\bf 29} 4337-4345 (1996)
%per le tre repliche  e per la quantita' tipo S-
%"Simulation of three-dimensional Ising spin glass model using three
%replicas: study of Binder cumulants"

\bibitem{CGGV} P. Contucci, C. Giardin\`a, C. Giberti, C. Vernia
{\em Phys. Rev. Lett.} {\bf  96}, 217204 (2006)


\bibitem{MaPa}
E. Marinari, G. Parisi
{\em Phys. Rev. Lett.} {\bf 86}, 3887-3890 (2001)
%per q link come funzione di q bisogna anche mettere
%"Effects of a Bulk Perturbation on the Ground State of 3D Ising Spin Glasses"



\bibitem{HYD} G. Hed,  A. P. Young, E. Domany
{\em Phys. Rev. Lett.} {\bf 92}, 157201 (2004)

\bibitem{CGGPV} P. Contucci, C. Giardin\`a, C. Giberti, G. Parisi, C. Vernia, in preparation.





\end{thebibliography}
\end{document}